**Math Agents: Computational Infrastructure, Mathematical Embedding, and Genomics**

Melanie Swan,[a] Takashi Kido,[b] Eric Roland,[c] Renato P. dos Santos[d]

[a]DIYgenomics.org; University College London (Research Associate)
[b]Advanced Comprehensive Research Organization, Teikyo University; Preferred Networks, Inc.
[c]RedBud AI, LLC
[d]Centre for Generative AI in Cognition and Education, Lutheran University of Brazil

**Abstract**
The innovation in generative AI could be further accelerated with more-readily usable and evaluable mathematics as part of the computational infrastructure. Beyond human-AI chat interaction, LLM (large language model)-based mathematical analysis tools are emerging in software programming, algorithm discovery, and automated theorem proving, but have not yet been widely applied to genomics. Towards disease-solving, this work introduces Math Agents and the mathematical embedding (vector-space representation of an equation as a data string) as new "Moore's Law of Mathematics" entries. The project consists of a GPT-based workflow to extract equations from published literature PDFs with Mathpix OCR and process them into LaTeX and Python embeddings. There are many ways to *represent* equations digitally, but few automated means for *evaluating* large bodies of equations (mathematical ecologies/mathscapes).

The important result of LLMs is that they are a linguistic user interface, a language-based access tool, via natural language for human-AI chat, but more extensively, via formal languages for at-scale AI-aided build-out of the computational infrastructure. AI tools are suggested as although the possibility space of natural language is relatively finite, formal possibility spaces are infinite (e.g. the programmatic space of algorithms, the mathematics space of theorems, and the computational complexity space of quantum-classical-relativistic classes).

Whereas humans interact with natural language, Math Agents interact with math, the implication of which could be a shift from "big data" to "big math" as a higher-order lever for interacting with reality. Natural language as a language is flexible and open to contextual interpretation; mathematics as a language has well-formedness properties subject to proof. Hence, mathematical use cases beyond math-as-math could include high-validation math-certified icons (by analogy to green seals) towards AI alignment aims of serving humanity in the broadest possible ways.

The current project develops a theoretical model for the deployment of Math Agents and mathematical embeddings to the information systems biology problem of aging, applying multiscalar physics mathematics (elucidating near-far entropic correlations in systems) to disease model mathematics and whole-human genomic data. Generative AI with episodic memory (per file dating/time-stamping) could assess causal relations in longitudinal personal health dossiers, deployed via SIR (sustaining, intervening, recovering) compartmental Precision Health models. In the short term, genomic variant and expression data is indicated for practical application to the unresolved challenge of Alzheimer's disease as the top-five human pathology with no survivors.

## Section 1: Introduction to Formalization Space

The contemporary moment is one of building heightened classes of digital infrastructure, in the form of smart network technologies running terrestrially and beyond which include AI, machine learning, blockchains, and quantum computing. A key emergence is AI language graphs, LLMs (large language models) – computerized language models generated with artificial neural networks (deep learning) which have billions of parameters and are pre-trained on large data corpora such as GPT-4 (OpenAI), LaMDA (Google), and LLaMA (Meta AI). In addition to a potential large-scale reorientation of white-collar labor with upleveled digital methods, an immediate conceptual result of LLMs is the first-principles thinking shift from "some" to "all" in terms of the level of consideration of the possibility space. It is increasingly routine to think of the entirety of a corpus such as *the* language space (of all human language), *the* program space (of all possible software programs), and *the* mathematics space (of all possible mathematics) (Figure 1). It is therefore obvious yet non-trivial, the idea of having a digitized and accessible mathematical space with easy-solve interfaces for a wider deployment of mathematics.

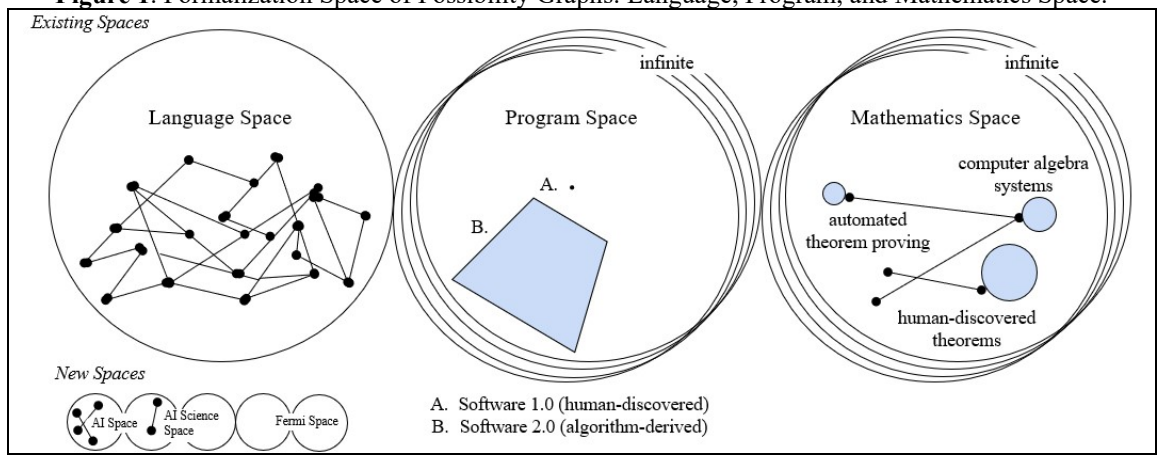

**Figure 1**. Formalization Space of Possibility Graphs: Language, Program, and Mathematics Space.

Other existing spaces could be added to Figure 1 such as computational complexity (the classes of resources needed to compute a problem), as well as emerging spaces including the AI space (of all AIs), the AI-driven Science Projects space, and more distantly, the Fermi (or Planck) space (of all foundational time and space models), to characterize the entries and relational activity in a possibility space. One high-order use of AI could be to elucidate and catalog formal spaces, as already demonstrated in the software program discovery space (Karpathy, 2017). Although some spaces are naturally formal (software programs, mathematics, computational complexity), others become formal (natural language) in the digitization process. "Formal" means a rigorous mathematical approach, but generally refers to the systematic consideration of a topic. The formalization space is the possibility space of all formal (systematic) approaches such as mathematical, algorithmic, programmatic, information-theoretic, and graph-theoretic methods. The computational infrastructure is the realization of formal methods on global networks (AI agents, machine learning, blockchains, and quantum computing). As data corpora (e.g. natural language, mathematics) are digitized, they join the computational infrastructure, becoming open to wide-spread democratized network access by all agents (human, AI, hybrid).

The benefit of LLMs is providing a linguistic interface overlay to the digital computational infrastructure. The interface allows humans to speak natural language and interact with the computational infrastructure (for a host of uses including but also going beyond language generation in the Siri-Alexa-chatGPT value chain). Of even greater potential impact is the possibility for AIs to "speak" any digitized formal language (natural language, mathematical symbols, programmatic code) to interact with and extend the computational infrastructure. This work proposes AI Math Agents and the mathematical embedding as new potential entries in the "Moore's Law of Mathematics" to facilitate the build out of the mathematical infrastructure.

Whereas the spaces of software programs and mathematics are lesser known and possibly infinite, the language space is familiar and (relatively) operationally finite. Before LLMs, the status was that although the structure of all human language was "known" in the sense of having been elucidated by scholarly efforts, this knowledge was not previously accessible as it was isolated in the brains of practitioners and academic researchers, published literature, and the earlier-stage digital method of language-learning applications (e.g. Duolingo, Babbel, Memrise). The new way of treating the entirety of a possibility space as a digital corpus means that language can be accessed globally with the click of a button. The benefit (though also risk) revealed by LLMs is the mobilization (democratized access) of freely-available digital corpora, for use by all agents. A similar trajectory for mathematics is proposed here.

Possibility space graphs raise the issue of novelty and reality status. Latency (existing but not manifest) is implied in the sense that all possible mappings pre-exist in the graph, even if not yet traced – articulated or accessed – by human or AI agents (Cheong, 2023). Every utterance pre-exists as a path in the graph. Voicing a new statement is merely walking a path through the language graph. A question as to what is truly new or original arises. In LLMs, at the level of form, the structure of all human language is known (as distilled from online data corpora). At the level of content, not all possible utterances have been said, but their form is likely to follow the recognized structure if they are to be intelligible. In one sense, the possibility space contains all currently existing form and content (subject to the latest LLM updates). In another sense, one of the reasons to encode digital possibility spaces is to facilitate "new utterances" (innovation) – to fill in gaps in the graph (Math Agents) and to push boundaries into new territories (human creativity, scientific discovery). As novelty occurs, it is incorporated into the graph accordingly.

*Mathematics as the Data Corpus*
In the contemporary project of the digitization of all data corpora (the world's currently-existing knowledge), mathematics is in the process of being digitally instantiated as any other data corpus. The entirety of mathematics (thus-far discovered though possibly-arbitrarily transcribed) is the data corpus that can be digitized and AI-learned for insight. Although considerable bodies of mathematics have been illuminated by a few centuries of human effort, such mathematics may be incomplete and could be integrated and extended with AI methods. Mathematics is widely regarded as being useful, but is generally not usable except by specialists, and while mathematics can be *represented* digitally with existing methods, it cannot be easily *solved*. This implies two points about the representation and evaluation aspects of the mathematical possibility space.

First, regarding representation (the easily-accessible digital representation of mathematics), a digital library and comprehensive catalog of the existing mathematics corpus is lacking. There are many ways to enter equations in a digital format, but not to call existing bodies of mathematical equations for easy deployment (like callable JavaScript libraries or a Wikipedia for mathematics). The overall shape and size of the mathematical space could be elaborated. It is not known but imprecisely estimated that a few million theorems have been human-discovered in the potentially infinite possibility space (Cottrell, 2012). The same mathematics may have been identified in different fields (e.g. eigenvalues (allowable scale tiers)) which could be linked or consolidated. Related bodies of mathematics might be integrated once seen in a comprehensive view. Computer algebra systems are one of the first steps towards a digital mathematical infrastructure. The deployment of newly-available AI methods is indicated as in terms of scale, some of the largest formal (computational) proofs are estimated to entail a human-unmanageable number of lemmas (Kepler conjecture initial proof (Hales, 2005) and formal proof (Hales et al., 2017); five-coloring problem (finding a Schur Number Five number of 160) with a two-petabyte sized proof (Heule, 2018)). Mathematics could join many areas of scientific discovery in which manual methods are giving way to AI-facilitated high-throughput digital computational methods. For example, the use of LLM tools for automated theorem proving (Yang et al., 2023).

Second, regarding evaluation, also at present, the mathematical data corpus is under-digitized and not readily usable and solvable. Mathematics is represented physically in hardcopy textbooks and digitally in LaTeX and PDF, and beyond basic education, engaged on a largely individual basis. There is no good way for sharing workflows and final mathematical ecologies produced in computer algebra systems (e.g. MatLab, Mathematica, Maple, SageMath). What is needed is a GitHub for mathematics. Although mathematics is represented digitally (primarily for publication), there are few tools for evaluating mathematics, particularly large mathematical ecologies in a systematic high-throughput manner. It is comparatively easy to represent mathematics, but quite difficult to automatically evaluate large bodies of mathematics. Mobilizing the useability of the mathematical corpus could render mathematics accessible and widely-usable as a tool, at different levels of deployment for different user audiences (in the analogy of HMTL pages, like a Geocities and Dreamweaver levels for mathematics).

*Mathematical Embedding*
An embedding is the numerical vector representation of a string, particularly to translate between high dimensional and low dimensional spaces. An initial step in machine learning systems is converting input data (whatever form of image, sound, text, medical results) to computer-readable strings (letters and numbers). In a standard example, the entries in the MNIST database of handwritten digits of "0" and "1" are converted to represent each image as a string of 784 numbers (each digit corresponding to the light-darkness ratio of the pixel value at that location in a 28x28 grid imposed on the image). The data strings (embeddings) are then fed into the input layer of a machine learning network for computational analysis.

Although embedding is used routinely in machine learning, the technique has mainly targeted the traditional content types of text, images, and sound. Hence, this work introduces the idea of the mathematical embedding as the similar mapping and instantiation of mathematical symbols and equations as strings of embeddings. Standard classes of mathematical embeddings could be callable in the computational infrastructure any time mathematical formulations are needed. Embeddings are not uniform due to parameter selections (i.e. for normalization, dimensionality reduction (UMAP, PCA, t-SNE), and tokenizing (data subset size-parsing (7x7 grid; 28x28 grid))), but could be packaged in templated versions (by analogy to Creative Commons licenses).

Word embeddings are widely used in LLMs in the context of news, social media, and online corpora such as Wikipedia, and are starting to be developed for scientific content (Figure 2). Three projects provide "possibility space" embeddings of an entire academic literature as the data corpus: (2a) all papers published in 2017 by journal (Meijer et al., 2021), (2b) all arXiv paper titles (2.3 million) (Depue, 2023), and (2c) a scalable interactive visualization tool of the ACL Anthology (Association of Computational Linguistics) (85,000 papers) (Wang et al., 2023). Another project (2d) generates embeddings for disease priority by gene, from the KEGG Pathway and Human Phenotype Ontology databases as an input to CNN machine learning (Li & Gao, 2019). The interpretation of all graphs is that clustering produces a human-readable signal.

**Figure 2**. Examples of Embedding Visualizations: (a)-(c) Academic Papers and (d) Disease Priorities by Gene.

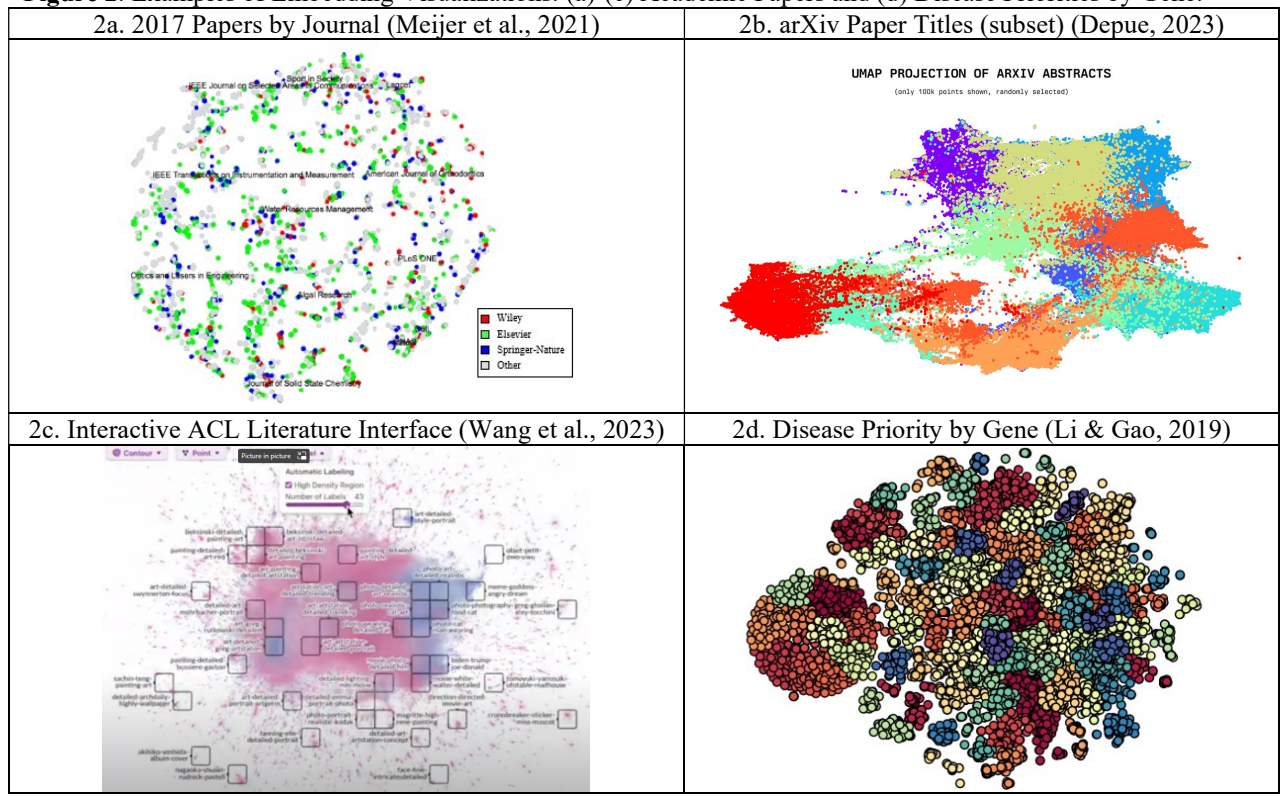

The current project introduces the mathematical embedding implemented with a GPT-based workflow to extract equations from published literature PDFs with Mathpix OCR and process them into LaTeX and Python embeddings. The mathematical embeddings are intended to cast mathematical equations into AI-readable format for further input to machine learning systems and the computational infrastructure. Mathematics as usable mobile units of digital abstraction could facilitate representation and evaluation of mathematical ecologies by humans and AIs.

*Math Agents*
The Math Agent is conceived as a learning-agent actor for orchestrating the digital mathematical infrastructure. An AI agent more generally is an artificial intelligence-based entity (set of algorithms) tasked with learning and problem-solving in a particular context, with feedback loops from the environment, per a rewards-driven action-taking policy. The Math Agent is the idea of AI agents as the operating interface on the digital mathematical infrastructure, acting to identify, catalog, represent, analyze, integrate, write, discover, solve, theorem-prove, steward, and maintain mathematical ecologies. Math Agents could help to further populate and integrate the mathematical possibility space and solve vast classes of mathscapes towards knowledge generation. Math Agents ("mathGPT") could serve as the dialogical interface between humans and the mathematical corpus to extend its use to Geocities-Dreamweaver user classes. Math

Agent functionality is already starting to be visible in augmenting the digital mathematical infrastructure, facilitating tasks such as computer algebra system automation, algorithm discovery (classical and quantum), and automated theorem proving (Yang et al., 2023).

The ability of AI to understand mathematical structures in ways that are not as readily accessible to humans connotes the ability to possibly make unprecedented progress in elaborating the mathematics of various large-scale multi-faceted human-facing situations such as clean energy, disease pathologies, and space exploration. The risk is also substantial, highlighting the growing power differential between human and AI capabilities, to which the present solution is AI alignment with RLHF (reinforcement learning with human feedback) and HITL (human in the loop) approaches at present to validate and censor AI output.

**Section 2: The Mathematical Embedding: Representation and Interpretation**
*Methods*
A standard machine learning method, vector embedding, is employed with mathematical equations and genomic data as the input. Vector embedding is the algorithmic processing of data into character strings for high dimensional analysis which is carried out and then translated back into low dimensional (2D) output for interpretation. In the mathematical embedding, equations are produced as vector embeddings in LaTeX (Figure 3) and SymPy (symbolic Python) (Figure 4). The result is that the whole of a mathematical ecology (set of equations) in a paper may be seen in one visualization. Further, the embedding visualization allows the comparative viewing of similar mathematical ecologies (Figure 7) as well as the math and the data together in one picture to assess correspondence between descriptive mathematics and underlying data (Figure 8). The mathematical embedding is at the equation-level, but could also be executed at the symbol-level, for "predict-next-symbol" in equation analysis, similar to predicting next word, phoneme, or letter in word-based LLMs. The mathematical embedding visualization of a 476-equation AdS/CFT correspondence mathematical ecology (Kaplan, 2016) is rendered with four standard mathematics-amenable embedding models (OpenAI, MultiQA, CodeSearch, and MathBert) in LaTeX (Figure 3) and symbolic Python code (Figure 4).

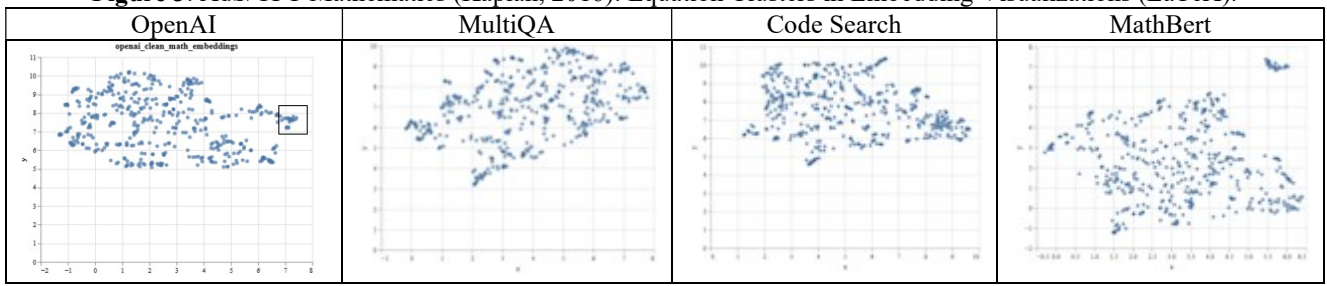
**Figure 3**. AdS/CFT Mathematics (Kaplan, 2016): Equation Clusters in Embedding Visualizations (LaTeX).

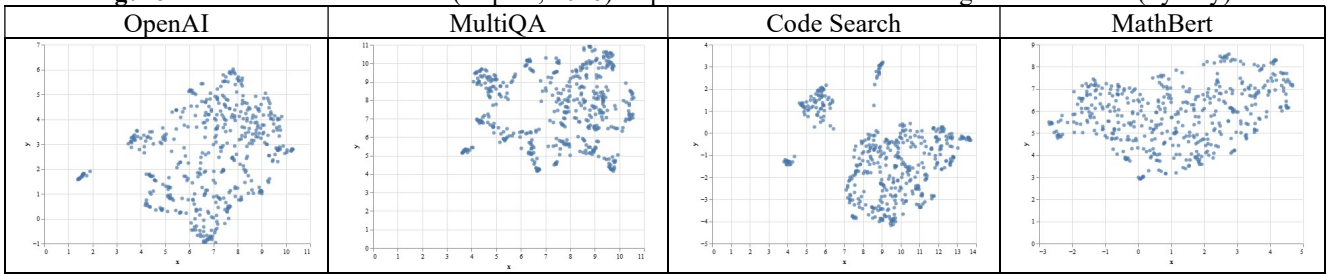
**Figure 4**. AdS/CFT Mathematics (Kaplan, 2016): Equation Clusters in Embedding Visualizations (SymPy).

Interpretatively, there is definite structure to each graph, suggesting embedding-identified interrelation between the 476 equations, realized slightly differently in each embedding model. The cluster concentration and use of white space indicates the grouping of similar kinds of equations, irrespective of their order of appearance in the linear progression of the paper. The units of the x and y axes (including negative values) are related to the embedding model-posted output to the graph and do not have human-interpretive value. More important is the relative clustering of the points (the embeddings are most conceptually similar to kernel learning methods). The Python code versions (Figure 4) show heightened equation clustering as compared with the LaTeX versions (Figure 3) mainly reflecting how the two representations are themselves different code-based languages of mathematics, both rendering mathematics in a usable digital format. LaTeX embedding visualizations are portrayed in this analysis bit for the next phase of equation evaluation, the Python code is implicated as being more mobile in readily joining the computational infrastructure. The OpenAI embedding model is also used as the standard in subsequent figures appearing in this analysis. Annotated views of the OpenAI model appear in Figure 5, illustrating (a) how the embedding method groups similar kinds of equations, and (b) the mouse-over view of equations annotated by equation number.

**Figure 5**. Annotated Equation Clusters in Embedding Visualizations: (a) Grouping-view and (b) Equation-view.

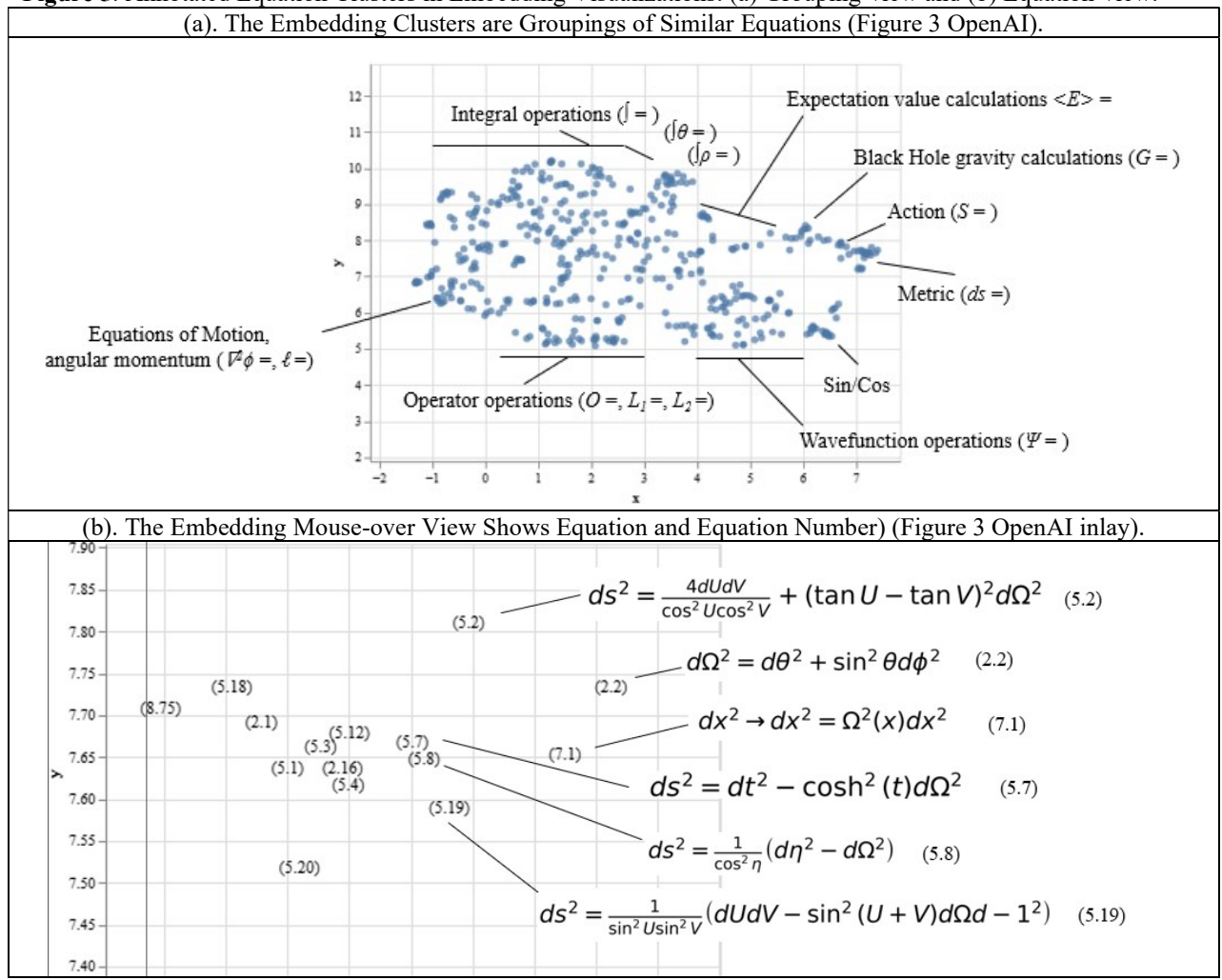

(a). The Embedding Clusters are Groupings of Similar Equations (Figure 3 OpenAI).

(b). The Embedding Mouse-over View Shows Equation and Equation Number) (Figure 3 OpenAI inlay).

*Data Selection*

This pilot project demonstrates the mathematical embedding as a step towards the realization of the AdS/Brain research program (applying mathematical physics to the biopathology of aging). A total of 10,592 embeddings were created for math and data (Figure 6). Embeddings were produced for the equations in the mathematical ecologies of 10 selected papers, and for the entirety of genomic variants (377 total, 276 unique) that have been associated with Alzheimer's disease, Parkinson's disease, and ALS (amyotrophic lateral sclerosis) in GWAS and EWAS studies (genome-wide and epigenome-wide association studies; underlying risk + expressed transcripts). An earlier project phase extracted the SNPs from research papers with GPT. Embeddings were generated for each equation and each RSID (genomic variant or SNP (e.g. rs6504163)) based on gene, effect, study size, and p-value. Eight embeddings were produced for each data element (equation or RSID), one for each embedding model (OpenAI, MultiQA, CodeSearch, MathBert) in LaTeX and SymPy (symbolic Python); OpenAI and LaTeX versions are reported in the results.

**Figure 6**. Embeddings Inventory.

|   | Category | Reference | Embeddings |
|---|---|---|---|
|   | *Mathematics-Physics* |   |   |
| 1 | AdS/CFT Correspondence | Hashimoto 2021 | 43 |
| 2 | AdS/CFT Correspondence | Guo 2016 | 68 |
| 3 | AdS/CFT Correspondence | Kaplan 2016 | 476 |
| 4 | Chern-Simons: DNA-RNA host-virus | Capozziello 2018 | 58 |
|   | *Mathematics-Disease Modeling* |   |   |
| 5 | Alzheimer's disease: transposon dynamics | Banuelos-Sindi 2018 | 44 |
| 6 | Alzheimer's disease: multiscalar aggregates | Kuhn-Sindi 2019 | 20 |
| 7 | Alzheimer's disease: tau phosphorylation | Hao 2016 | 20 |
| 8 | Alzheimer's disease: protein kinetics | Fornari 2020 | 86 |
| 9 | Alzheimer's disease: protein clearance | Thompson 2021 | 94 |
| 10 | SIR Compartmental Model (control) | Wyss 2023 | 38 |
|   | **Total Mathematical Equations** |   | **947** |
|   | *Genomic Data* |   |   |
| 11 | Alzheimer's, Parkinson's & ALS GWAS-EWAS | Various (15 papers) | 377 |
|   | **Total SNPs (RSIDs)** |   | **377** |
|   | **Total Embeddings: OpenAI** |   | **1,324** |
|   | **Total Embeddings: OpenAI, MultiQA, CodeSearch, MathBert** | (x4) | **5,296** |
|   | **Total Embeddings: LaTeX and SymPy** | (x2) | **10,592** |

GWAS-EWAS (genome-wide and epigenome-wide association studies)

*Mathematical Embedding Workflow*

To treat mathematical equations as the data corpus, a novel AI-driven machine learning workflow was developed. Equations were identified and extracted from PDF papers with OCR into Mathpix images and then produced as LaTeX and SymPy (Symbolic Python) embeddings using mathematics-conducive standard embedding models (OpenAI, MultiQA, CodeSearch, and MathBert). GPT was used as a core computational component in the data processing workflow to validate mathematical output and embeddings. Other similar workflows for equation digitization are starting to be proposed (Eskildsen, 2023).

UMAP (uniform manifold approximation and projection) was selected as the dimensionality reduction method. This is because UMAP has more mathematical well-formedness features for local-global analysis in systems than the other options, t-SNE (t-distributed stochastic neighbor embedding) and PCA (principal components analysis). UMAP incorporates the global structure of the data by preserving the broad layout of points in addition to small-scale relationships. The local-global structural features of the mathematical embedding suggest a parallel embodiment to the physics mathematics for modeling holographic and information-theoretic entropy formulations in the analysis of near-far correlations in systems such as transposable element dynamics influencing insertion-deletion activity in the genome.

*Results*

In the project, 10,592 total embeddings were produced from the mathscapes in 10 papers and 276 unique genomic variants implicated as being associated with Alzheimer's Disease. The data corpus is 4 multiscalar mathematics physics papers (3 AdS/CFT correspondence papers and 1 Chern-Simons theory paper), 5 Alzheimer's disease mathematics papers, and one control example, with SIR compartmental model mathematics. The 3 AdS/CFT papers were selected out of a population of 30 identified AdS/CFT mathematics papers as those elaborating a specific "Applied AdS/CFT" use case. The Chern-Simons paper applies Chern-Simons theory to DNA-RNA host-virus interaction which is relevant as host-virus interaction is likewise implicated in triggering transposable element movement in Alzheimer's genomics. The 5 Alzheimer's disease mathematics papers selected are all of those with a substantial identifiable mathematical model.

**Figure 7**. Equation Clusters in Mathematical Embeddings (a) AdS/CFT and Chern-Simons (b) Alzheimer's Disease.

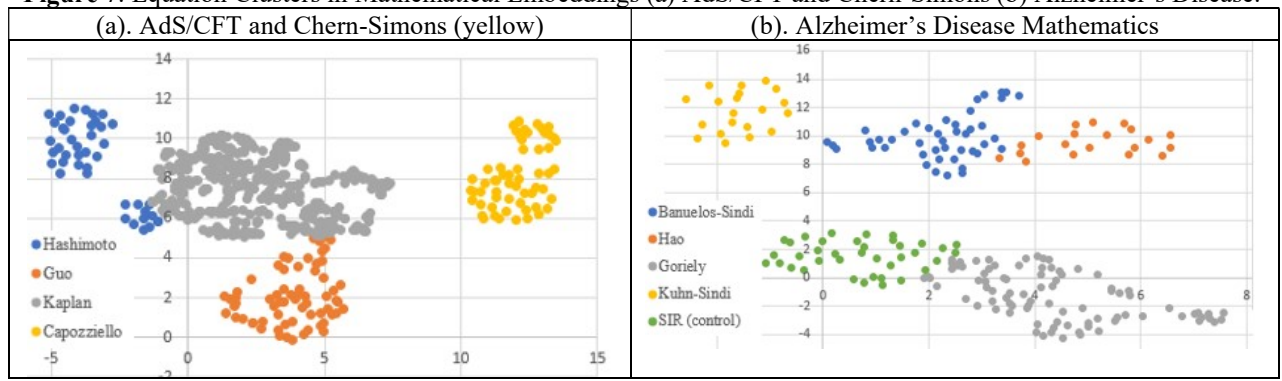

The mathematical embedding provides the ability to view the entirety of a mathscape (set of equations) in a paper in one at-a-glance abstracted and consolidated view, and compare related mathematical ecologies. Figure 7a portrays the 3 AdS/CFT mathematical ecologies and the Chern-Simons mathematical ecology. On the one hand, it is surprising that there is not more overlap in the AdS/CFT mathematical ecologies since they are ostensibly of the same mathematics. On the other hand, this is higher math, and the mathscapes of the 43, 68, and 476 equations, respectively, are elaborated separately for different purposes. The AdS/CFT mathematics is a somewhat standard mathematics which equates a messy bulk volume with a boundary field theory in one fewer dimensions. The mathematics has the core structure of four equations defining the structure of the bulk space (the Metric ($ds =$)), operators to act on the bulk and the boundary (Operators ($O =$)), the system dynamics from which equations of motion can be derived (the Action ($S =$)), and the system energy operator (the Hamiltonian ($H =$ )). The hypothesis would be that producing embeddings for the full set of 30 AdS/CFT mathematics ecologies might reflect more overlap, at least in the core equations that state the correspondence. From a math studies perspective, the equation clusters and embeddings suggest that it will not necessarily be straightforward to integrate mathematical ecologies.

In Figure 7b with the Alzheimer's disease mathematics, the lack of overlap is not surprising as the programs target different phenomenon: transposon dynamics, multiscalar aggregates, tau phosphorylation, and protein kinetics and clearance. The reason that the SIR control model and the Goriely mathematics are close together is because both have a heavy focus on differential equations in their mathscapes. The Alzheimer's disease mathematical ecologies help as a first-pass view of the landscape of the math, without having to read and digest the paper, with the easy mouse-over view to see the types of mathematics used by the authors in the form of the equation clusters. One research question is whether the Alzheimer's disease mathematical ecologies should be connected, whether that would be helpful in obtaining a causal understanding of the pathology. There is also the possibility of adding a time dimension to the Equation Cluster embeddings as different mathematics may describe the different phases of the pathology's evolution (the data for 3 phases of Alzheimer's onset and progression now exists, and these data could be accompanied by the relevant mathematics).

**Figure 8**. Equation Clusters and Data Embedding Visualization: (a) Transposon Math-Data and (b) AdS Math-Data.

| (a). Transposon Dynamics + Chern-Simons + AD RSIDs | (b). AdS Mathematics + AD RSIDs |
|---|---|

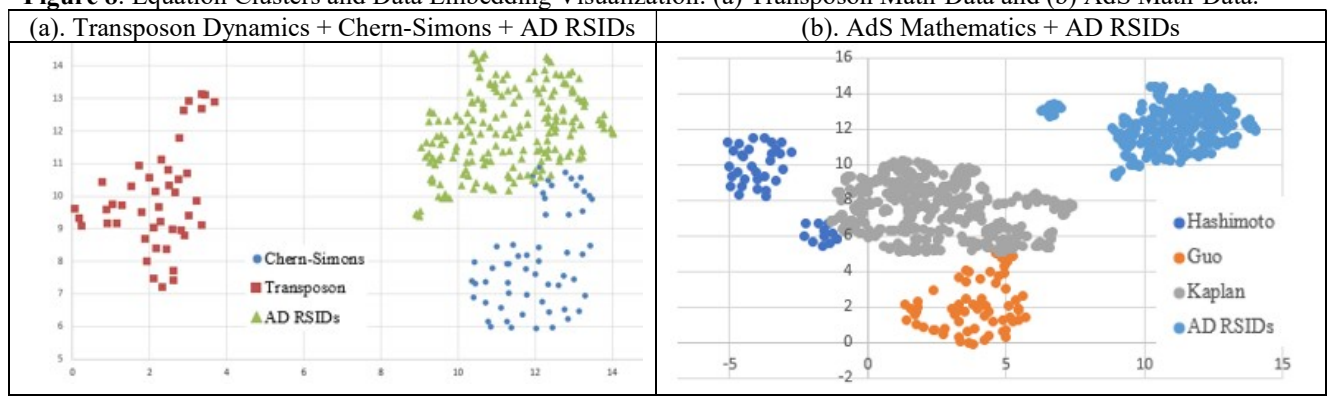

The charts in Figure 8 join mathematics and data in the same view with embedding as the common vernacular. Figure 8a shows the mathematical ecologies for Chern-Simons DNA-RNA host-virus interaction, transposon dynamics, and Alzheimer's disease SNPs, and Figure 8b shows the mathematical ecologies for AdS/CFT and Alzheimer's SNPs. The reason for Figure 8a is to investigate the transposon claim of Alzheimer's disease genesis, which is that viral infection leads to transposable element insertion-deletion movement in DNA which triggers Alzheimer's disease onset. The equation clusters are of two possible math ecologies that might explain this claim, which, when viewed together with the data, suggest that the Chern-Simons model may have higher explanatory value as the embedding clusters are in closer proximity to those of the data. However, this is only a first-pass hypothesis using the mathematical embedding abstraction tools, as direction-setting for further investigation. In Figure 8b, equation clusters for the AdS/CFT mathematics appear together with the Alzheimer's disease SNPs. It is not readily human-discernable how these data sets may relate, except for possibly that the Kaplan math is closer to the Alzheimer's data. The method, if any, for working with a mathematical ecology and an underlying data set in a unified embedding visualization is not yet clear. The premise here is that it may be useful to look at the math and data of a problem together through the same lens, that of the embedding, which provides a meta level for examining math and data simultaneously. Even if the two are not related, it is useful to see the size and shape of a mathematical body and a data body in the embedding format.

**Figure 9**. (a). Embeddings Data View: Alzheimer's SNPs and (b-c) Precision Medicine Citizen 1, 2 AD SNPs.

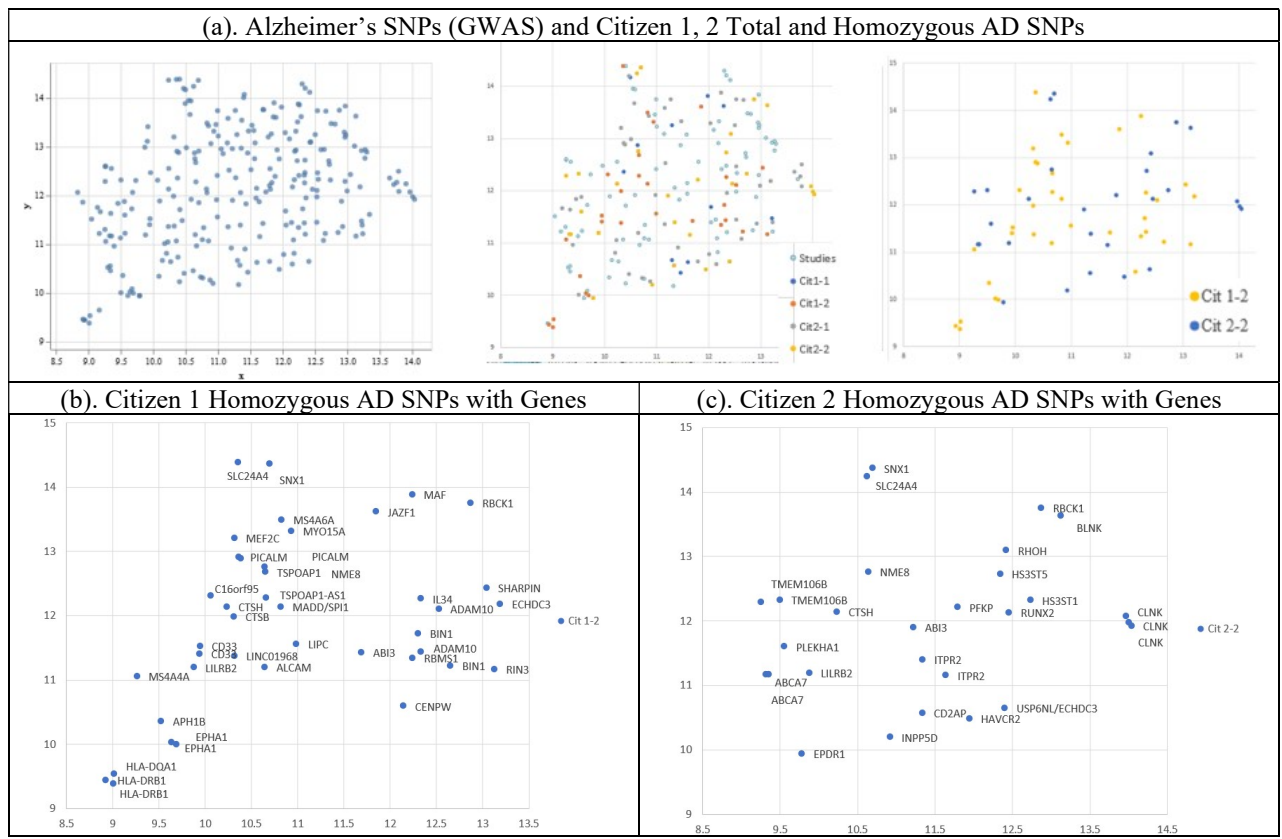

Figure 9 provides a data-only view of genomic variants and how they might be linked to Precision Medicine initiatives. Figure 9a is the embedding of all SNPs (RSIDs) associated with Alzheimer's disease large-scale association studies. Figure 9b-c is from the total variants, those SNPs for which the whole-human genome analysis of two individuals, Citizen 1 and Citizen 2, indicates homozygosity (two alternative alleles) and therefore higher risk or interventional starting points. Notably, each individual is homozygous for different subsets of genes; Citizen 1 for more immune system related SNPs such as HLA-DRB1 and CD33, and Alzheimer's-related clathrin binder (PICALM). Citizen 2 is homozygous for more membrane proteins and cytokine-dependent hematopoietic cell linkers (CLNK). Both are homozygous for the solute carrier protein (SLC24A4) and nexin (SNX1).

This could be actionable information as genomic medicine shifts into a new and vastly more expansive era requiring AI assistance to analyze epigenetic markers, transposable element insertions-deletions, and eQTLs (expression quantitative trait loci), the genomic directions to express harmful proteins (a locus that explains a fraction of the genetic variance of a gene expression phenotype). The point of Figure 9 is to show how these kinds of AI embeddings approaches might be helpful in the realization of precision health initiatives, identifying specific genes, pathways, expression profiles, and blood biomarkers, and their causal interactions, that are medically actionable.

The applied genomics precision health implication is starting with known condition alternative allele SNPs in the healthy citizen patient, and obtaining eQTL RNA expression and blood plasma data as needed in causality-tracking personal health dossiers. AI methods engaging mathematics and data are also implicated in broader information system biology approaches to link various diseases previously considered on a standalone basis such as Alzheimer's disease and diabetes (Alzheimer's disease as a form of Type 3 Diabetes), ApoE-based links between Alzheimer's disease and Down's syndrome (Dooling et al., 2022) and the potential interrelation between Alzheimer's disease, Parkinson's disease, and ALS (amyotrophic lateral sclerosis) as the embedding visualization of the overlap in genomic SNPs as a starting point (Figure 10b).

*AdS/Brain Research Program*
The AdS/Brain research program develops the idea that multiscalar mathematical models discovered in physics (such as the AdS/CFT correspondence and Chern-Simons topological invariance) may be relevant in theorizing and modeling the multiscalar complexity of biology. This could be useful for information systems biology problems such as aging and chronic disease for which a causal understanding is lacking. Leading multiscalar mathematics programs (the AdS/CFT correspondence and Chern-Simons theory) as mathematical ecologies (sets of equations) may be applied to model the complexities of biosystems which are likewise rigorously multiscalar. The AdS/Brain theory has been proposed theoretically but not tested experimentally, which may start to be feasible with AI-driven at-scale science methods (Swan et al., 2022a,b). The aim of the current project is to explore AI-related tools in an integrated approach to solving disease pathology that has not been easily available prior to AI methods, in the proximate investigation of Alzheimer's disease. The hypothesis is that multiscalar physics mathematics (elucidating near-far relations in systems) may be applied to develop a causal understanding of the pathologies of aging in a genomic theory of medicine involving SNP variations (single nucleotide polymorphism) and transposable element dynamics. Genomics is indicated in Alzheimer's disease activation (variants linked to disease risk and expression (Sun et al., 2023)), and LLMs are seen as being an indispensable at-scale tool for genomic analysis (Nguyen et al., 2023; Batzoglou, 2023). Into this trajectory, the current work formulates a mathematically-based approach to Alzheimer's genomics.

*Control Example: Two-tier Precision Medicine SIR Model: Health and Information Flow*
A control example is selected for this analysis to provide a comparison between the current work and a familiar widely-known example of mathematics from epidemiology (Figure 10a). This is the SIR compartmental model (all individuals in a population are in the categories of being either susceptible, infected, or recovering), expressed as a set of solvable differential equations (dS/dt, dI/dt, dR/dt) (Kermack & McKendrick, 1991). The equation clusters in the mathematical ecology visualization confirm the presence of the expected mathematics, and show how the embedding organizes the mathematics into different equation clusters for the system's differential equations both together (SW corner) and separately (NE and S), and other clusters for parameter specification (NE corner). The SIR model is not only useful for comparison as a known mathematics, but also in formulating a mathematical approach to realizing precision medicine.

As a compartmental model of the population, the SIR model may be adapted to the precision medicine use case of individuals being in a new cast of SIR states, those of sustaining, intervening, or restoring. An important feature of the traditional SIR model is tracking infectious disease spread, however, in the precision health model, "infection" is the "positive infection" of the spread of information which impels preventive intervention. There could be a two-tier model, the underlying tier of the health graph of individuals in SIR states, and the secondary tier of the information flow of inputs from physicians, apps, clinical trials, health research studies, advocacy groups, health social networks, virtual patient modeling, and quantified-self citizen science efforts to implement necessary preventive initiatives. The aim of precision medicine is to prevent conditions early in the 80% of their lifecycle before they become clinically detectable. A Precision Health SIR Model could be a formal way to implement precision health initiatives in society, possibly in concert with other quantitative models such as value-based healthcare (compensation tied to health outcomes).

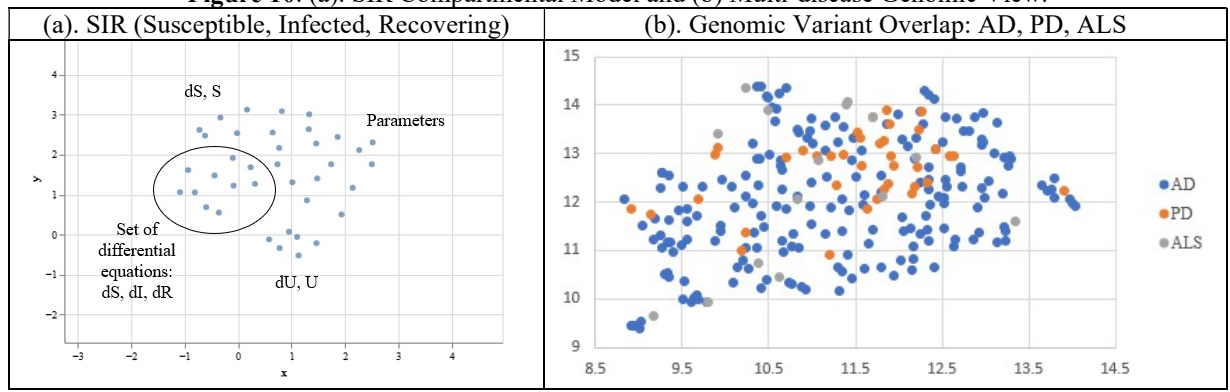

**Figure 10**. (a). SIR Compartmental Model and (b) Multi-disease Genomic View.

*Interpretation: Mathematical Position Evaluation*
The point of the current pilot project is to develop a suite of AI-based mathematical abstraction tools that might be helpful towards in the construction of the digital mathematical infrastructure. The result is producing several high-level tools for viewing and working with mathematical abstraction, together with data sets, namely, the mathematical embedding, equation cluster, and mathematical ecology visualization, as new forms of high-level mathematical abstraction tools. From this demonstration, the broader use case, if any, for the deployment of these tools is not yet fully clear, from the perspective of both human and AI agents. However, these kinds of high-level abstraction tools, used in concert with Math Agent explorers might contribute to math discovery as new elements in the mathematical infrastructure for various operations, including position evaluation – being able to quickly grasp the size and scope of a mathematical ecology.

The intended audience of embeddings is AI systems; embeddings render input in AI-readable form for further AI-based analysis. However, embeddings are also useful to human audiences in providing the ability to see the entirety of a mathscape consolidated in one view with zoomable levels of abstraction, and zoomable in-out capability to examine different aspects of the math. Vector embeddings themselves are emerging as one element in the larger standard body of computational infrastructure in which AI agents can help tackle systems-level problems too complex to address with previous methods. Although embeddings are developed in a human-orchestrated workflow here, they are increasingly being automatically incorporated into the cognitive infrastructure in new waves of GPT release. The kinds of mathematical signals expressed in the new infrastructure have a proximate use in being mobilized in the next steps of building large-scale precision health models to identify and prevent disease.

**Section 3. Math-Data Relation**
AI tools offer more expedient ways to interact with reality at the level of mathematics instead of data. Such a trend is ongoing as the human interaction with the computational infrastructure is already one of mathematics on the computer side as user interfaces employ quantitative formal methods to produce the experience which humans receive as qualitative, with all the rich data properties of everyday existence ("Alexa, turn on some music").

What is different is an upleveling in the concept of what the digital infrastructure is, and the kinds of interactions it affords for humans, possible shifting from a "big data" to "big math" era. Whereas the past mode of interaction with the digital infrastructure focused on corralling "dumb" data, there is an awareness of the digital infrastructure now having "smart" (self-computational, formal, math-y) properties, e.g. automatic spell-checker, word-completion, recommendations, advertising-serving. This contributes to the sense that "the network is alive," which AI copilot technologies in web interfaces and document processing further accentuate.

One research question is how AI tools may be changing the math-data relation. The math-data relation is the correspondence between mathematics and data. The embedding is an interesting technology in that it renders mathematics and data in the same format. On the one hand, math and data can be analyzed in the same view. On the other hand, math and data are two different things, and it may be misleading to examine them in the same view. Progressive ideas of how the math-data relation may be changing are presented in Figure 11 as the math-data composite view of two representations of the same system, the multiscalar renormalization of viewing a system at any scale tier through the lens of a conserved quantity, and mathematical abstraction as the interface (Kantian goggles) for data sublation and human and AI interaction with reality.

**Figure 11**. Mathematics as Foundational Lever for Interacting with Reality: The Math-Data Relation.

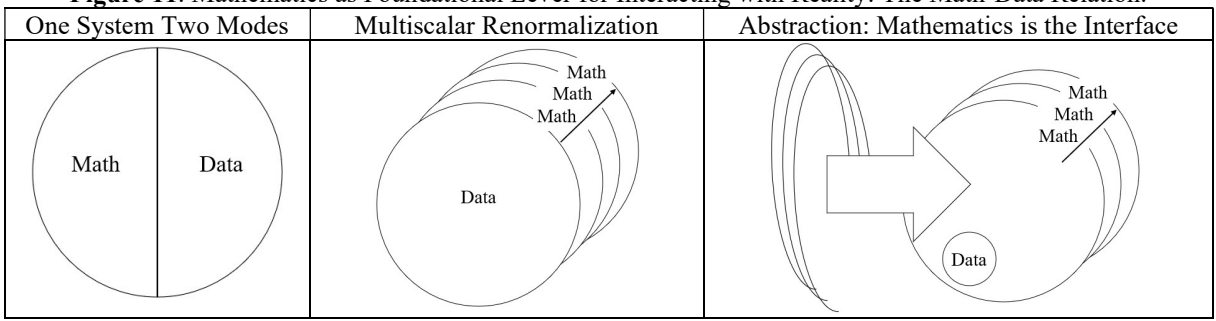

*One System Two Modes*
The math-data relation often centers around the model-fit problem. The model-fit (map-territory) problem is how well a generalized set of descriptive mathematics accurately recapitulates the data of an underlying phenomenon in which exceptions, variation, and irregularity may be inherent. The trade-off is how much specificity may be lost to capture the core salience of the phenomenon. By example, diagnostic healthcare is a notoriously high-dimensional realm in which there are many approaches to the model-fit problem (Montoya & Edwards, 2020).

To the extent that it is valid and useful to examine the mathematics and the data of a system through the same lens of embedding (the mathematical embedding and the data embedding), model-fit problems may be better assessed. Theoretically, the math and the data are two representations of the same system, and one could be viewed from the other (if the math is accurate). For example, there could be a set of data points for how many sales occurred each day, and an accompanying mathematical curve fit through the data which allows an estimate of future sales by day (absent holidays and other irregularities). The "sell more widgets" system can be examined with either lens, as data points or as a mathematical curve. Seeing math and data together suggests the idea of a math-data composite view.

The math-data composite view is seeing the math and the data for a system together in one view, which is demonstrated in this analysis through the embedding, but there could be other methods. Manipulable zoomable 3D visualization tools, data explorers and math explorers, could allow the size and shape of a "data set of data" and a "data set of math" to be viewed, both separately and superimposed so their correspondence may be assessed, by humans and AI Math Agents. Just as a curve's model-fit through a set of data points can be somewhat readily interpreted, the idea is a more complicated setup in which different mathematical ecologies applied to different layers of complex data sets may be assessed, to determine the model-fit between a certain math and data.

In application to science, although a lot of mathematics have been proposed, there is little confirmation, validation, mobilization, replicability, and ease of deployment. Given scale limitations, some large portion of the previous scientific endeavor may have been that of a specialized mathematics developed to model a specific small set of data. However, the at-scale full possibility space thinking enabled by contemporary tools means that the scientific endeavor can be recast to include the targeting of comprehensive solutions and understandings of vast multiscalar ecosystems and the mathematics that models and causally and predictively explains them. Such comprehensive knowledge at scale is the aim of AI systems and tools such as the Math Agent and the mathematical embedding. At minimum, these advances may offer new ways to consider the data stack, math stack, and math-data relation at multiple scale tiers in complex systems together with various actor-observers (humans, AIs, biological processes) interacting with the system in different ways (Figure 12). There could be advantages to working math as an entity, data as its own entity, as well as the data-math composite as a new kind of formal entity.

**Figure 12**. Math, Data, Math-Data Stacks and Scale-Tier Actor-Observers in Genomic Medicine.

| Scale | Data Stack | | Math Stack | | Math-Data Stack | |
|---|---|---|---|---|---|---|
| Macro | Actor -> | Genetic variants | Actor -> | Signal processing math | Actor -> | Genetic variant near-far relations |
| … | Actor -> | | Actor -> | | Actor -> | |
| Meso | Actor -> | Transposon indels | Actor -> | Topological biomath | Actor -> | Transposon dynamics |
| … | Actor -> | | Actor -> | | Actor -> | |
| Micro | Actor -> | Amyloid-beta plaque, tau tangles | Actor -> | Biochemical math | Actor -> | Protein biomarker distribution |

*Multiscalar System Mathematics*
One way to conceive the math-data relation is as two sides of one coin or as two different representations of the same system. Another way to see the math-data relation is at different levels of complexity, for example, that the data is an imperfect messy bulk volume and the math is a streamlined equation on the boundary describing the bulk volume. Emergent structure in bulk data can be detected by boundary math. This is the premise of machine learning: write a function describing these data. Since a multiscalar system is connected, every scale tier is an upleveled abstraction from the detail that takes place at the tiers below. For example, a gecko produces a limb as an end product, with many specific underlying steps and coordination. One elaboration of such causal activity between tiers in biosystems is denoted "multiscale competency" in which (in regenesis and development) any scale tier appears to call the entire functionality of lower scale tiers (Levin, 2023).

Particularly in biosystems, there are a number of intricate unpredictable relations between scale tiers so the challenge is not as simple as aggregating up to "temperature" from "particles." In considering the human brain and its pathologies, nine order-of-magnitude scale tiers have been identified (Sejnowski, 2020) involving 82 billion neurons, 242 trillion synapses, and neuron-glia interactions (Martins et al., 2019). By analogy to economics, the math and the data for the micro and macro scale tiers are known (individual transactions and GDP), but not a full and systematic model of causal behavior in the middle tiers.

Hence, the treatment of complex systems entails not only cataloging scale tiers, but the causal interaction between levels and how the entity operates together as a system. The open nature of biological systems is also important as actor-observers interact in constant feedback loops with the environment. Given these specificities, it is not possible to simply apply normalization (typically percentage-based scaling around 1 or 100) as a formal method to put scale tiers into dialogue. This immediately suggests the physics-based multiscalar method of renormalization. Renormalization is a formal method which allows the ability to view a system at multiple scales by collapsing parameters (degrees of freedom) that are not relevant across scale tiers. Instead, a system-wide factor may be identifiable such as symmetry (in the universe) or free energy (in biological systems) that is conserved, represented, and engageable at different scale tiers.

Some of the best-known renormalization programs for viewing systems across difference scale tiers are AdS/CFT and Chern-Simons. The holographic correspondence, AdS/CFT (anti-de Sitter space/conformal field theory), refers to the possibility of describing a messy bulk volume with a boundary theory in one fewer dimensions, any physical system, whether the universe, a brain, a bug on a windshield, or a room (Maldacena, 1999). Chern-Simons theory is a model of topological invariance ("bending not breaking" unchanging aspects of a system as other changes are applied), a solvable quantum field theory in which (nonlocal) observable measures (Wilson loops) can be represented as knots (that generalize to a known knot invariant, the Jones polynomial) (Chern & Simons, 1974). Both AdS/CFT and Chern-Simons apply across all physical scales (quantum-classical-relativistic).

*Abstraction: Mathematics is the Interface*
As a formal data corpus, mathematics is contiguous in ways that other data corpora are not. This suggests that to some extent, even the simplest equation calls the entire corpus of existing and possible mathematics as there may be arbitrarily-many layers of subsequent abstraction (e.g. set theory, category theory, type theory). The idea of mathematical complexity connotes that all mathematics at a certain level of abstraction may be computationally-equivalent in requiring the same degree of computational resources to solve. From a practical perspective, the interconnectedness of mathematics implies a framework for identifying the right level at which to solve multiscalar systems. The implication is working smart not hard. In the protein folding example, both nature and AlphaFold do not try every possible permutation but smart-solve more directly to the final conformation, using chemical bonding energy cues (nature) and high-level algorithms (AlphaFold (Jumper et al., 2021)).

The bigger implication of a transition to digitized mathematics is the ability to interact with reality at a higher level. Mathematics could provide a more effective and efficient means of interacting with reality. This is not a new thought. Mathematics is central to the scientific method. The aim is to write descriptive mathematics of a phenomenon such that new predictions about future behavior can be made. Mathematics is a content with high truth value. As more aspects (complex systems and everyday phenomena) of reality become expressible in mathematics, a more effective lever is available for engaging them. Multiple levels of human interfaces to the digital mathematics corpus are required, for example, at the levels of professional mathematician, scientific practitioner, economist, marketing, and lay persons.

The mathematical picture provides a different kind of Kantian goggles as a perceptual interface to reality, not just a quantitative scale revealing of the big and the small with the telescope and the microscope, but qualitative access to a more foundational structure of reality. Just as solving a multiscalar system at the right tier is important, so too, upleveling the overall interaction with reality at the right structural level might be similarly right-sized. The further implication of interacting with reality at the level of mathematics is a potential shift from the "big data" era to the "big math" era. Once data are corralled into the digital infrastructure with automated methods (a non-trivial task), mathematics as an abstracted level at which to interact with data is implied. This could be the new math-data relation, the idea of big data -> big math in treating digital reality at higher levels of abstraction which conferring greater effectiveness of salience, relevance, efficiency, and results.

**Section 4: Evaluation of Mathematical Ecologies**
Mathematical embeddings or some other form of mathematics as mobile units of digital abstraction are implicated not only for representing large systems of equations (mathematical ecologies) but also for solving them. The implication of mathematics instantiated in a usable graph-based digital architecture is potential for the automated evaluation of mathematical ecologies. Mathematical embeddings are vector-space strings which are in AI-readable format as input to machine learning systems that run on graphs. Just as any equation joins the mathematical corpus of abstraction, so too any graph-formulated entity joins the entirety of the graph-theoretic smart network infrastructure which includes machine learning, blockchains, and quantum computing. Graphs facilitate access to the full range of information-theoretic properties such as energy-entropy formulations, uncertainty relations (quantum scale), and statistical interpretation. The digitized format suggests that automated Math Agent-facilitated evaluation could proceed at a number of levels in the math stack ranging from the brute-force testing of all permutations to the application of various higher-level mathematical methods to more efficiently solve a system.

The result of mathematics in the form of graphs is that the corpus interfaces well and can be taken up into the existing smart network infrastructure, which is likewise instantiated in graphs (machine learning, blockchains, quantum computing), in an overall picture of the computational infrastructure of digital reality. Easy-use human interfaces are indicated. The digitization of possibility spaces (data corpora) includes meta-level tools for large-scale mobilization such as "top-level kernels" calling the entirety of the corpus and time-stamping clocks for cause-and-effect tracking. The language of formal methods is abstraction, deployed as Merkle root hashes in blockchains and embeddings in machine learning.

The digital approach to mathematics (with AI tools such as mathematical embeddings, equation clusters, Math Agents) is as follows. Mathematics is instantiated in graphs: a mathematical ecology (equations or symbols) is instantiated as the nodes of a graph. Solving proceeds in lock-step movement through the graph, with Math Agents finding the best path, by analogy to blockchain path-routing. The smart network graph is a math engine with Math Agents running on it. Mathematical embeddings and Math Agents that walk on these graphs to evaluate mathematical problems serve as two new elements in the digital mathematical infrastructure that have easy-to-use dialogical interfaces which offer the ability to interact with reality at a higher level. The representation and evaluation of mathematics may be incorporated in any variety of smart network technologies such as machine learning, blockchains, and quantum computing, as well as the network itself ("the network is the computer") (Figure 13). The idea is Math Agents (AI bots) running on the possibility space of the mathematical graph, finding new solution paths in the mathematical possibility space. The result could be a continuing transition from the "big data" era to the "big math" era. Math Agents are stateful information states with action-taking policies, running as a smart graph overlay, taking actions to proceed through the graph as a state-based information system to find new solutions.

Figure 13. Computational Infrastructure. Mathematical Functionality and Smart Network Technology.

|   | Smart Network Technology | Mathematical Functionality: Representation and Evaluation |
|---|---|---|
| 1 | Machine learning | Transformer evaluation of entire mathematical corpus at once |
| 2 | Blockchain | Equation evaluation, theorem-proving, and IP-logged discovery (mNFT) |
| 3 | Quantum computing | Uncertainty relation entropy-energy trade-offs (ID cross layer correlations in multiscalar systems); renormalization AdS/CFT DMRG-Q |
| 4 | The network is the computer | Network-informed information states (computation outsourced to network) |

mNFT: mathematics NFT (non-fungible token): blockchain-registered mathematical entity (theorem, proof, lemma)

*Machine Learning*
Machine learning is the marquis digital infrastructural technology, now in routine use in nearly all fields. Embeddings are the standard means of data input to machine learning systems, as any kind of data is encoded as vector-space strings and passed into the graph. The contemporary machine learning model used by GPT and other LLMs is transformer neural networks, which offer an advance by analyzing all data simultaneously to find relations between tokens (small packages of data). In the main machine learning method, problems are formulated as optimizations, and the network cycles (forward and backpropagating) to find the best weightings for network nodes to deliver a predictive solution. Physics and economics principles inform machine learning in the use of lowest-energy cost functions that descend a gradient to identify the best solution. In the widely-used Boltzmann machine method, an energy-minimizing probability function is used to evaluate machine learning algorithm output (based on the Boltzmann distribution from statistical mechanics in which the probability that a system will be in a certain state is evaluated as a function of the state's energy and system's temperature). Math Agents could run as an overlay to machine learning to evaluate mathematical ecologies, including incorporating abstraction in different ways.

The most straightforward implementation is mathematics being analyzed as the data corpus in the usual machine learning network setup. A secondary implementation could involve the mathematical problem structure being setup in the machine learning network architecture for the system to find the best weights, number of network layers, and overall graph shape corresponding to the mathematical solution. By analogy, a real-world physics problem (the emerging bulk structure of a quark-gluon plasma) is set up in a holographic machine learning model in which the emerging neural network structure corresponds to an emerging physical bulk structure that describes chiral condensates (Hashimoto, 2021). The point is that machine learning networks incorporate a variety of graph-theoretic features that could be employed by Math Agents to evaluate mathematical ecologies, namely, properties such as physics-inspired energy-entropy loss function calculations, causal inference, and probabilistic prediction.

*Blockchains*
Blockchains (distributed ledger systems) are likewise a smart network technology with properties conducive to the Math Agent evaluation of mathematical ecologies, especially with the Layer 2 abstraction layer. The conceptual idea of a blockchain in this context is a living graph that can do math. Whereas both machine learning and blockchains are smart network systems in which problems are instantiated in graphs, blockchains provide more robust functionality. Machine learning graphs operate for high throughput and efficiency with minimal operations at each node, namely the up-down weighting of a probability coefficient as the network cycles backward and forward to obtain an optimal predictive result. Blockchain graphs offer richer node functionality for mathematical equation evaluation and proof. These kinds of smart network features could be useful when high-throughput data processing is not the aim but rather the creation of a self-solving digital mathematical infrastructure. Blockchains, for example, are core providers of clocks (dos Santos, 2019), time-keeping mechanisms in the computational infrastructure which could be incorporated into AI episodic memories to potentially attribute causality to event unfolding such as pathology development (Shapiro, 2021).

Solving math in blockchain networks could proceed as follows. Layer 2 overlays such as the Lightning Network run as an abstraction layer to blockchains, with useful functionality for blockchain operates such as allowing a batch of administrative transactions to be consolidated and posted to the blockchain, automated operations such as wallet balancing, smart contract execution, oracle confirmation, and transaction path routing. Nodes are stateful and agents can engage these states in audit-logged transaction paths for mathematics use cases such as equation evaluation and automated proof generation. The Layer-2 overlay enables path routing through the graph in which nodes are equations or lemmas, through which a math agent cannot proceed without appropriate validation (confirmation of the validity of the mathematics at this node).

First considering equation evaluation, the mathematical ecology proposed in a published paper (~50-200 equations on average) can be instantiated in a math graph for evaluation, validation, and replication. Each equation is a node in the math graph (other embeddings could be at the level of symbol or token). The Math Agent proceeds through the math graph nodes to confirm and evaluate the chain of equations. Blockchain functionality tracks and administers this process, confirming progress at each node with hash functions before progressing to a subsequent node is allowed. The AI features of the Math Agent can help to address the problem of replicability (that some high percent of mathematical equation ecologies in published literature cannot be simply implemented and run as a whole). In its math graph validation processes, the Math Agent may be able to solve in-situ, filling in clarificatory adjustments or additional mathematics to required proceed through the math graph with the result of producing a tested robust generalized version of the mathematical ecology. Blockchain math agent certifications (with a $F(x)$ symbol for example) could reflect the fact that a body of mathematics has been independently tested and certified. Any new proposed mathematics (whether human, AI, or human-AI discovered) could attest to its validity by seeking independent Math Agent certification. In the overall frame of integration, synthesis, and extension of the mathematical corpus, routine tasks for the Math Agents could include finding and applying the best math for a data set, assessing model-fit between math and data, and evaluating multiple mathematical ecologies as an ensemble. To solve a problem, Math Agents might deploy any of the mathematics at their disposal from the digital library of the mathematical infrastructure including machine learning technique such as genetic algorithms, as they solve at different levels of abstraction or computational complexity in the mathematical methods stack.

Second, the Math Agent blockchain infrastructure can be used for mathematical proofs. Automated theorem proving is already part of the digital mathematical infrastructure and could be further extended with blockchain features. As with path-routed evaluation, a blockchain-based Math Agent cannot advance to the next node (equation) in a mathematical ecology without confirming the current node, and proof technology layers more functionality onto this. The blockchain step-by-step retraceable transaction log is already conducive to the likewise step-by-step structure of theorem proving. In a more formal blockchain mathematical proof structure, the Math Agent could execute a transaction at each node confirming the validation path, transactionally enacted with an administrative allotment of MathProofCoin.

The blockchain audit log is useful not only to document the mathematical proof (in the step-by-step format of proofs), but also for a host of related administrative functions. These could include the automatic minting of new NFTs for proofs in IP (intellectual property) discovery blockchains (MathChains, both MathProofChains and MathTheoremChains), and the logging of unique identifiers in a theorem-addressing system to facilitate finding and calling such theorems later in the digital library of the mathematical infrastructure. The ownership of such proofs might reside with the Math Agent discovering them. The role of MathProofCoin as a token has further potential uses in credit-assignment tracking of how contributed theorems are used (similar to program code use tracking in subsequent software applications), and as a resource allocation mechanism for Math Agent renumeration and as a contribution barometer for further resources to be allocated to the development of needed and promising mathematical ecologies.

Economics is a set of principles for interacting in graphs. Aside from conducting proofs on blockchains with Math Agents, there are other related benefits of managing the AI Agent ecosystem with blockchains. The proliferation of AI entities suggests AI registries to track behavior, development, and societal impact, including from a legal, regulatory, and liability perspective. AI Agents could be registered as operating entities in AI Blockchain Registries. There could be AI-related analogs to GAAP (Generally-Accepted Accounting Principles) in the form of GAAiP (Generally-Accepted AI Principles), with a framework of reporting requirements and annual audit overseen by the "FINRA" (Financial Industry Regulatory Authority) of AI, "FAiNRA." AI registries as an element of the computational infrastructure can be envisioned to orchestrate AI entities with verified identity and liability accountability. Lawsuits involving AI entities can be imagined: who is liable – the Cigna AI virtual patient modeling bot? As engineers sign bridges and bioengineers sign synthetic biology creations, AI products could be tracked likewise, with a variety of AI certifications in the vein of CC-Licenses for AI ethics adherence and behavioral conduct. Tokens allocated to registered AI entities, AIcoin, could be used for administrative matters such as the cost of registration and compliance (to fund FAiNRA) as well as multiagent coordination. Blockchain consensus generation among agents could proceed with a peer mining model (transaction confirmation is a utility function not a wealth-generation financial incentive as any would-be network user must confirms two other transactions (randomly-assigned) before being able to submit their own transaction to the network. AIcoin could be instantiated with lifecycle management demurrage-type principles so that currency allotments expire on an annual basis, thus constructing the token as a purely administrative mechanism disallowing accretion and economic control.

*Quantum Mathematical Infrastructure*
The fact that machine learning models already incorporate physics-based energy principles is conducive to a potential AI-quantum computing convergence. Lowest-energy formulations are important in both machine learning and quantum mechanics, and in addition, energy and entropy are related terms. In classical machine learning in the Boltzmann machine model, algorithm-generated output is evaluated based on loss functions, gradient descent, and lowest-energy probability. In the Born machine as the quantum analog, probabilistic quantum system output is similarly evaluated with the Born rule (the probability density of finding a particle at a given point, calculated by squaring wavefunction amplitudes).

Energy and entropy are related in that entropy refers to the number of possible microstates of a system and measures how much energy has been dispersed in a process. Since energy flows from high to low, entropy tends to increase with time (absent the injection of new energy, a desk becomes messier over time not cleaner). In information theory, entropy indicates the number of bits (qubits) required to send a message given some error rate, as Shannon entropy in the classical setting, and von Neumann entropy (taking the minimum over all measurement bases of Shannon entropy) and Rényi entropy (the generalization of Shannon, Hartley, collision, and minimum entropy) in the quantum setting. The Heisenberg uncertainty relation (trade-offs between partner properties such as position-momentum and time-energy) is also employed operationally as an efficient means of calculating entropy (Yunger et al., 2019).

Embedding in the generic mathematical sense is used in quantum physics to isolate or freeze parts of a system to facilitate calculations in the overall system. The core quantum object, the density matrix (all the information of a quantum state), is treated through the density matrix embedding theory (many-body embedding of arbitrary fragments of a quantum system) and the density matrix renormalization group (finding the lowest-energy Hamiltonian in a system). In one project, a team embeds the DMRG algorithm in an environment to solve quantum chemistry problems (Dresselhaus et al., 2014). In another project, the team couples a polarized embedding approach to the DMRG to select a region within a larger quantum system while still modeling the surrounding environment. (Hedegard & Reiher, 2016)

These ideas are relevant in the sense that AI tools such as Math Agents are implicated in integrating the classical and quantum infrastructure together in a potential era of quantum computing. As part of the digital infrastructure, Math Agents may be able to operate more translationally than humans between classical, quantum, and relativistic domains. Like other smart network setups, quantum computational problems are solved in graphs. Energy-entropy formulations provide a higher-level yet still well-formed (tied to math and physics) level of abstraction for the expedient solving of a system at the right scale tier. Renormalization techniques may be applied to the graph itself, as relating energy in the graph is similar to relating free energy across tiers in a biosystem, or symmetry across tiers in the universe. Energy-entropy formulations could become the abstraction layer at which graph problems are solved. As the Boltzmann machine centers on the lowest-energy gradient descent, so too quantum graph technologies solve for the lowest-energy activity level in the graph. The blockchain math agent evaluating equations, proving theorems, and discovering new algorithms can path-route through the lowest-energy configuration of the graph, or conversely, when the objective is the opposite structure, the highest-entropy configuration of the graph (in quantum-secure lattice-based cryptography). The idea is having digital infrastructure that incorporates natural energy-based formulations to a greater extent (not just Feynman's computing with atomic object but also principles of atomic energy). In practice, the Math Agent could evaluate equations as the lowest-energy path through the math space. In other future work, there could be more extensive non-classical formulations of embeddings in hyperbolic vector space, and with non-linear time.

*The Network is the Computer*

The "network is the computer" scenario is the natural progression endpoint in which active cognitive agent functionality is incorporated into network itself so that *network-informed information states* drive the infrastructure. It would be indistinguishable whether there is an agent running on the network or if it is an agent-enabled network. The Math Agent as an actor-observer in the cognitive architecture could operate as an overlay to networks or within networks. Instead of AI agents running on the network engaging data corpora (math, language, images), the cognitive functionality could be implemented directly into the network itself.

The more speculative farther future could be one in which the network is alive, in this sense, with mathematics, with computing; the network is the computer; the network is the math. The level of the computational stack at which the Math Agent is deployed is less relevant that the role it fulfills in validating, affirming, safeguarding the content in various digital possibility spaces and generating, integrating, extending the knowledge base of how mathematical ecologies are related to one another as a core digital mathematical infrastructural tool.

From the network point of view, in the network-eye view, the network is the observer-agent. Traditionally the network can interact with data but not math; in the network agent frame, the network can also interact with math (or any formal data corpora). Computation could be outsourced to the network for certain operations instead of the traditional edict of keeping the network exclusively for transport. The network could be reconceived as not simply a dumb transport layer, but rather itself as the location of richer levels of computation. Instead of computation only taking place at network ends for transport efficiency, there could be some still-efficient smart processing taking place in the network itself, as smart infrastructure layers. This would be within the standard 7-layer OSI network stack which includes abstraction layers: physical, data link, network, transport, session, presentation and application, which are morphing to accommodate the cognitive infrastructure of AI technologies (AI chips) and quantum networks (entanglement heralding, quantum key distribution)

"Math networks" could be more extensively developed as a smart network layer with mathematical well-formedness functionality. Higher mathematical formulations could be incorporated into the computational infrastructure in the concept of "math networks" as a math network layer which includes AI-readable mathematical embeddings orchestrated, operated, and maintained by Math Agents. Math networks could join other "network is a living graph" ideas such as network-informed information states that enable self-propagated action-taking by other network-based agent-entities. The Math Agent and network-enabled AI functionality are examples of the emerging cognitive architecture (Shapiro, 2021) in that Math Agents (AI bots) could run as actor-observer operators on the digital mathematical smart network infrastructure, finding new paths through the graph of the mathematical possibility space.

**Risks and Limitations**
There are many different risks and limitations associated with this work. First and foremost is at the level of AI in that any AI project must consider AI Alignment, taking all steps to produce AI that has broadly humanity-benefiting values. The current work enacts this ethos by being pointed at widespread democratized access to mathematics and disease-preventing healthy well-being. This work follows and endorses initiatives underway to establish global AI regulatory agencies with innovation sandboxes, registries, and accountability frameworks (FLI, 2023). Within this context, the current project aims to produce easy-to-use accessible interfaces to mathematics as a digital data corpus for the implementation of these tools to solve global health problems in delivering high quality of life and equitable economic impact. Other work discusses the Math Agent and human-AI entities in the responsible development of quantum intelligence (AI-enabled technologies in quantum-classical-relativistic domains) (Swan & dos Santos, 2023).

Second, at the level of the current project, there are issues related to the proposal of a digital mathematical infrastructure and AI-enabled tools such as Math Agents, equation clusters, and mathematical embeddings. The first point is that AI functionality is constantly evolving, and any solution may face "immediate obsolescence." However, this does not invalidate the need to create prototypes that harness AI technologies in new ways, including for human comfort and understanding. This project demonstrates the mathematical embedding in theory and practice, as a potential precursor step to embeddings starting to be included as standard functionality in AI engines such as GPT. The second point is the claim that just because math is produced in consumable digital units does not make it any less recalcitrant to the general human understanding. However, the goal is to render mathematics human-deployable without having to be human-understandable, in the analogy of an automobile (trusted operation without detailed knowledge). At present, mathematics is a specialist-access only data corpus, but could be a general-access tool, as it occurs to more endusers that "There's a math for that!"

**Conclusion**
The current moment may be one of a period of accelerated computational infrastructure build-out, the audience of whom is AI first and primarily and humans second. The surprise is that LLM functionality is for AIs, not for humans, serving as a linguistic interface for AI access to all digitally-rendered languages including human natural language, programmatic code, and mathematics. The research question is how AI agents may be deployed to elucidate these formal spaces and discover new ones, all towards broadly humanity-serving aims. The implication is AI-based tool deployment to hasten scientific discovery in biology, energy, and space science, for example in new cloud-based high-sensitivity use cases such as personal brain files.

Smart network infrastructure is becoming cognitive infrastructure – the network is the computer. Instead of smart technologies running *on* networks (AI, machine learning, blockchains, quantum computing), smart technologies are *becoming* the network; the network is becoming a cognitive agent. Whereas human observer-agents interact with data, AI observer-agents interact with math. Any observer-agent is simply accessing "reality" through whatever set of perceptually-constrained goggles they have, all of which are simply datapoints as each entity reaches out to describe "the elephant" of the encountered world. The idea of different user audiences accessing differentially-viewable formal spaces and technology-enabled levels in the same overall reality is explored in the concept of user-selected tech-locks (Schroeder, 2005), and also by Kant (1781).

We have long been using technology tools to extend human perception beyond the evolutionary Kantian goggles of sensory experience in 3D space and 1D time. Telescopes and microscopes reveal scales of reality that are not directly human-perceivable, and how matter properties are different in quantum mechanical and relativistic domains. There is an awareness that not only matter properties, but also time and space are fundamentally different in the "Planck space" of quantum mechanical and relativistic domains, properties which are necessarily incorporated into the modern computational infrastructure (e.g. quantum computing, GPS). In a Copernican shift, the familiar everyday 3D space and 1D time are unseated as the norm. There is a growing awareness that Kantian goggles are merely one actor-observer interface on a larger reality with different flavors of space and time; spherical-flat-hyperbolic space and linear time as a user-selected parsing mechanism through possible simultaneity.

Although humans have always been inventing various kinds of goggles to extend perceptual reach, the mathematical view provides a different kind of Kantian goggles as a perceptual interface to reality, not just a quantitative scale revealing of the big and the small with the telescope and the microscope, but qualitative access to a more foundational structure of reality. We do not human-see mathematics, but Math Agents could provide the interface. Mathematics is argued to be foundational (quark properties are quantitative; distinguished exclusively by numbers (mass, charge, and spin) (Tegmark, 2021)), but what is new is having a potential way to interact with reality at this level and test the claim. Math Agents could be Kantian goggles that extend human reach to other formal spaces (programmatic code, mathematics, computational complexity), and also possibly other forms of intelligence, artificial, computational, and quantum, and biological (overlooked due to lack of the right translational interface).

The result of this work is to introduce a suite of AI math tools – mathematical embeddings, equation clusters, mathematical ecologies and mathscapes, and AI Math Agents – to enable the further development of the digital mathematical infrastructure. This could also include the mathematics-certified symbol $F(x)$ as an icon to designate digital mathematically-verified content. The result of a cognitive infrastructure that includes mathematics in a more readily usable form is a modern advance which delivers democratized access to mathematics to a wide range of agential audiences (human and AI) at various theoretical and practical levels. Mathematics is widely regarded as a high-value discovery, but one that is under-developed, under-deployed, and perhaps near the limit non-AI-aided methods. A digital mathematical infrastructure may have uses not only in expanding mathematical discovery, deployment in a new larger slate of human-facing challenges in health, energy, and space, but also as a high-validation technology for achieving AI alignment goals with humanity-serving values and global health-preserving well-being.

**Supplemental Information**
*Software Code Availability*
Open-source Code Repository:
https://github.com/eric-roland/diygenomics

*Data Availability*
Whole-human Genome
Citizen 1: (Nebula):
https://www.diygenomics.org/citizengenomics/rsid_query.php (Citizen 17)
Citizen 2: (Illumina): http://startcodon.org/
https://www.diygenomics.org/citizengenomics/rsid_query.php (Citizen 18)

**Glossary**
**Agent**: AI (artificial intelligence) predictive entity (set of algorithms) tasked with learning, problem-solving, and behavior-updating in a context per a rewards-driven action-taking policy
**AI Alignment**: AI compatibility with human values (responsible broad humanity-serving values)
**Autonomous Cognitive Entity**: AI agent empowered to undertake autonomous operations
**Cognitive Infrastructure**: computational infrastructure that incorporates AI agent capabilities
**Computational Infrastructure**: concept of digital reality as a collection of formal methods

**Digital Mathematical Infrastructure**: mathematics digitized as an easy-access data corpus, learned and mobilized by Math Agents for insight and deployment operations
**Formal methods**: systematic mathematically rigorous techniques for operating in a context
**Formalization**: rendering a domain with formal methods (systematic mathematical techniques)
**Formalization space**: the possibility space of all formal (systematic) approaches: mathematical, algorithmic, programmatic, information-theoretic, graph-theoretic, computational complexity
**Embedding**: character string representation of a data element in high dimensional vector space
**Equation Cluster**: similar equations grouped in mathematical ecology embedding visualization
**HITL (human in the loop)**: AI results supervised and validated by a human
**Human-AI Entities**: RLHF partnerships operating as substrate-agnostic scale-free intelligence
**Math Agent**: AI agent operating in digital mathematical domain to identify, represent, analyze, integrate, write, discover, solve, theorem-prove, steward, and care-take mathematical ecologies
**Mathematical ecology (mathscape)**: set of related mathematical equations
**Mathematical embedding**: mathematical entity (symbol, equation) represented as a character string in vector space for high-dimensional analysis in AI-based machine learning systems
**RLHF (reinforcement learning human feedback)**: iterative dialogical human-AI interaction; AI agents that have a learned model of the environment, a decision-making policy, and a reward prediction mechanism, engaged in iterative feedback loops with humans
**Smart Network technologies**: AI agents, machine learning, blockchains, quantum computing available as standard formal methods deployed in the computational infrastructure

**References**
Banuelos, M. & Sindi, S. (2018). Modeling transposable element dynamics with fragmentation equations. Mathematical Biosciences. 302:46–66. https://doi.org/10.1016/j.mbs.2018.05.009.

Banwarth-Kuhn, M. & Sindi, S. (2019). Multi-Scale Mathematical Modeling of Prion Aggregate Dynamics and Phenotypes in Yeast Colonies. Biomathematics. 1-30. http://dx.doi.org/10.5772/intechopen.88575.

Batzoglou, S. (2023). Large Language Models in Molecular Biology: Deciphering the language of biology, from DNA to cells to human health. Towards Data Science. 2 June 20213. https://towardsdatascience.com/large-language-models-in-molecular-biology-9eb6b65d8a30.

Capozziello, S., Pincak, R., Kanjamapornkul, K., and Saridakis, E.N. (2018). The Chern-Simons current in systems of DNA-RNA transcriptions. Annalen der Physik. 530(4):1700271. doi:10.1002/andp.201700271.

Cheong, B. (2023). The Wilds of Artificial Intelligence. Entitled Opinions. https://entitled-opinions.com/2023/03/13/the-wilds-of-artificial-intelligence-with-bryan-cheong/.

Chern, S.-S. & Simons, J. (1974). Characteristic Forms and Geometric Invariants. Annals of Mathematics. Second Series. 99(1):48-69. https://doi.org/10.2307/1971013.

Cottrell, S.S. (2012). How Many Theorems Are There? Ask a Mathematician/Ask a Physicist. https://www.askamathematician.com/2012/11/q-how-many-theorems-are-there/.

Depue, W. (2023). Embeddings for every research paper on the arXiv. Twitter. 25 May 2023.

https://twitter.com/willdepue/status/1661781355452325889?lang=en. arXiv Title and Abstract embeddings available at: https://alex.macrocosm.so/download.

Dooling, B., Johnson, N.R., Joshi, M. et al. (2022). Interrogating the role of APOE4 in Alzheimer's disease and Down syndrome using human induced pluripotent stem cells (hiPSC)-derived cerebral organoids. Alzheimer's & Dementia. https://doi.org/10.1002/alz.068061.

Dos Santos, R.P. (2019). Consensus Algorithms: A Matter of Complexity? Swan, M., Potts, J., Takagi, S., Witte, F. & Tasca, P., Eds. Blockchain Economics: Implications of Distributed Ledgers - Markets, Communications Networks, and Algorithmic Reality. London: World Scientific. Pp. 147-170.

Dresselhaus, T., Neugebauer, J., Knecht, S. et al. (2014). Self-Consistent Embedding of Density-Matrix Renormalization Group Wavefunctions in a Density Functional Environment. arXiv:1409.1953v1.

Eskildsen, S. (2023). Parsing Math Equations. Twitter. 26 June 2023. https://twitter.com/Sirupsen/status/1673309920769323008?cxt=HHwWgMC9qbyg5bguAAAA.

Fornari, S., Schafer, A., Kuhl, E. & Goriely, A. (2020). Spatially-extended nucleation-aggregation-fragmentation models for the dynamics of prion-like neurodegenerative protein-spreading in the brain and its connectome. Journal of Theoretical Biology 486:110102. https://doi.org/10.1016/j.jtbi.2019.110102.

Future of Life Institute (FLI). (2023). Policymaking in the Pause What can policymakers do now to combat risks from advanced AI systems? 19 April 2023 https://futureoflife.org/wp-content/uploads/2023/04/FLI_Policymaking_In_The_Pause.pdf.

Guo, M., Niu, C., Tian, Y. & Zhang, H. (2016). Modave lectures on applied AdS/CFT with numerics. In: Proceedings, 11th Modave Summer School in Mathematical Physics. arXiv:1601.00257v2, PoS Modave2015(2016)003.

Hales, T.C. (2005). A proof of the Kepler conjecture. Annals of Mathematics. 2(162):1063–1183. https://annals.math.princeton.edu/wp-content/uploads/annals-v162-n3-p01.pdf.

Hales, T.C., Adams, M., Bauer, G. et al. (2017). A Formal Proof of the Kepler Conjecture. Forum of Mathematics, Pi. 5(e2). https://doi.org/10.1017/fmp.2017.1.

Hao, W. & Friedman, A. (2016). Mathematical model on Alzheimer's disease. BMC Syst Biol 10(108). https://doi.org/10.1186/s12918-016-0348-2.

Hashimoto, K., Hu, H.-Y. & You, Y.-Z. (2021). Neural ordinary differential equation and holographic quantum chromodynamics. Mach Learn: Sci. Technol. 2(03):035011.

Hedegard, E.D. & Reiher, M. (2016). Polarizable Embedding Density Matrix Renormalization Group. J. Chem. Theory Comput. 12(9):4242–4253. https://doi.org/10.1021/acs.jctc.6b00476.

Heule, M. (2018). Schur Number Five. Proceedings of the AAAI Conference on Artificial Intelligence. 32(1). https://doi.org/10.1609/aaai.v32i1.12209.

Jumper, J., Evans, R., Pritzel, A. et al. (2021). Highly accurate protein structure prediction with AlphaFold. Nature. 596:583–589. https://doi.org/10.1038/s41586-021-03819-2.

Kant, I. (1988 [1781/1787]). Critique of Pure Reason. Trans. & Ed. P. Guyer & A.W. Wood. Cambridge: Cambridge University Press.

Kaplan, J. (2016). Lectures on AdS/CFT from the bottom up. Johns Hopkins Lecture Course.

Karpathy, A. (2017). Software 2.0. Medium. 11 November 2017. https://karpathy.medium.com/software-2-0-a64152b37c35.

Kermack, W. & McKendrick, A (1991). Contributions to the mathematical theory of epidemics – I. Bulletin of Mathematical Biology. 53(1–2):33–55. doi:10.1007/BF02464423.

Krantz, S.G. (2007). *The Proof is in the Pudding: A Look at the Changing Nature of Mathematical Proof*. Cham Switzerland: Springer.

Levin, M. (2023). Darwin's agential materials: evolutionary implications of multiscale competency in developmental biology. Cell Mol Life Sci. 80(6):142. doi: 10.1007/s00018-023-04790-z.

Li, Y. & Gao, X. (2019). PGCN: Disease gene prioritization by disease and gene embedding through graph convolutional neural networks. bioRxiv: http://dx.doi.org/10.1101/532226.

Maldacena, J.M. (1999). The large N limit of superconformal field theories and supergravity. Intl. J. Theor. Phys. 38(4):1113–1133. doi:10.1023/A:1026654312961.

Martins, N.R.B., Angelica, A., Chakravarthy, K. et al. (2019). Human Brain/Cloud Interface. Front. Neurosci. 13(112):1–23. doi:10.3389/fnins.2019.00112.

Meijer, H.J., Truong, J. & Karimi, R. (2021). Document Embedding for Scientific Articles: Efficacy of Word Embeddings vs TFIDF. arXiv:2107.05151v1.

Nguyen, E., Poli, M., Faizi, F. et al. (2023). HyenaDNA: Long-Range Genomic Sequence Modeling at Single Nucleotide Resolution. arXiv:2306.15794v1.

Schroeder, K. (2005). Lady of Mazes. New York: Tor.

Sejnowski, T.J. (2020). The unreasonable effectiveness of deep learning in artificial intelligence. Proc. Natl. Acad. Sci. U.S.A. 117(48): 30033–30038. doi/10.1073/pnas.1907373117.

Shapiro, D. (2021). Natural Language Cognitive Architecture: A Prototype Artificial General Intelligence. https://github.com/daveshap/NaturalLanguageCognitiveArchitecture.

Steingart, A. (2022). *Axiomatics: Mathematical Thought and High Modernism*. Chicago: University Chicago Press.

Sun, N., Akay, L.A., Murdock, M.H. et al. (2023.) Single-nucleus multi-region transcriptomic analysis of brain vasculature in Alzheimer's disease. *Nat Neurosci.* 26:970–982. https://doi.org/10.1038/s41593-023-01334-3.

Swan, M. & dos Santos, R.P. (2023). Quantum Intelligence: Responsible Human-AI Entities. AAAI 2023 Spring Symposium: Socially Responsible AI for Well-being. South San Francisco CA, March 27–29, 2023. https://www.slideshare.net/lablogga/quantumintelligence-responsible-humanai-entities.

Swan, M., dos Santos, R.P., Lebedev, M.A. & Witte, F. (2022a). *Quantum Computing for the Brain*. London: World Scientific. https://doi.org/10.1142/q0313.

Swan, M., dos Santos, R.P. & Witte, F. (2022b). Quantum Neurobiology. *Quantum Reports*. 4(1):107–127. https://doi.org/10.3390/quantum4010008.

Tegmark, M. (2021). *Our Mathematical Universe: My Quest for the Ultimate Nature of Reality*. New York: Knopf.

Thompson, T.B., Meisl, G., Knowles, T.P.J. et al. (2021). The role of clearance mechanisms in the kinetics of pathological protein aggregation involved in neurodegenerative diseases. J. Chem. Phys. 154:125101. https://doi.org/10.1063/5.0031650.

Wang, Z.J., Hohman, F. & Chau, D.H. (2023). WIZMAP: Scalable Interactive Visualization for Exploring Large Machine Learning Embeddings. arXiv:2306.09328v1. https://github.com/poloclub/wizmap.

Wyss, A. & Hidalgo, A. (2023). Modeling COVID-19 Using a Modified SVIR Compartmental Model and LSTM-Estimated Parameters. Mathematics. 11:1436. https://doi.org/10.3390/math11061436.

Yang, K., Swope, A., Gu, A. et al. (2023). LeanDojo: Theorem Proving with Retrieval-Augmented Language Models. arXiv:2306.15626v1.

Yunger, N.H., Bartolotta, A. & Pollack, J. (2019). Entropic uncertainty relations for quantum information scrambling. Commun Phys 2(92). doi:10.1038/s42005-019-0179-8.